\documentclass[11pt,a4paper]{article}
\usepackage{jcappub}

\newcommand{\veps}{\varepsilon}
\newcommand{\pat}{\partial}

\title{Numerical analysis of cosmological models for accelerating Universe in
Poincar\'e gauge theory of gravity}
\author[a,c]{A.S. Garkun,}
\author[b]{V.I. Kudin,}
\author[c,d]{A.V. Minkevich}
\author[c]{and Yu.G. Vasilevsky}
\affiliation[a]{The National Academy of Sciences of Belarus} \affiliation[b]{Belarusian State
Technical University} \affiliation[c]{Belarusian State University} \affiliation[d]{Warmia and
Mazury University in  Olsztyn}

\emailAdd{garkun@bsu.by} \emailAdd{minkav@tut.by} \emailAdd{kudzin\_w@tut.by}

\abstract{Homogeneous isotropic models with two torsion functions built in the framework of the
Poincar\'e gauge theory of gravity based on general expression of gravitational Lagrangian by
certain restrictions on indefinite parameters are analyzed numerically. Special points of
cosmological solutions at asymptotics and conditions of their stability in dependence of indefinite
parameters are found. Procedure of numerical integration of the system of gravitational equations
at asymptotics is considered. Numerical solution for accelerating Universe without dark energy and
dark matter is obtained. It is shown that by certain restrictions on indefinite parameters obtained
cosmological solutions are in agreement with SNe Ia observational data and Big Bang Nucleosynthesis
predictions. Statefinder diagnostics is discussed in order to compare considered cosmological model
with other models.}


\begin{document}

\maketitle

\section{Introduction}

One of the most principal recent achievements of observational cosmology is the discovery of the
acceleration of cosmological expansion at present epoch \cite{riess98,perlmutter}. In order to
explain this observable accelerating cosmological expansion  in the framework  of General
Relativity Theory (GR), the notion of dark energy (or quintessence) was introduced in cosmology.
According to obtained estimations, approximately 70\% of energy in our Universe is related with
some hypothetical kind of gravitating matter with negative pressure --- ``dark energy'' --- of
unknown nature. Previously a number of investigations devoted to dark energy problem (DEP) were
carried out  (see reviews \cite{prev2,prev3}).

According to widely known opinion, the dark energy is associated with cosmological term, which is
related in the framework of standard $\Lambda CDM$-model to the vacuum energy density of quantized
matter fields. In connection with this the following question appears: why the value  of
cosmological term is very small and  close to average energy density in the Universe at present
epoch (see for example \cite{prev4}). Other treatment to solve the DEP is connected with
modification of gravitation theory.

One of such solutions can be obtained in the framework of gravitation theory in
the Riemann-Cartan spacetime $U_4$ - Poincar\'e gauge theory of gravity (PGTG)
\cite{kibble,brodskii,sciama,hehl1,hayashi}. It should be noted that the PGTG
is a natural and in certain sense necessary generalization of metric theory of
gravitation by including the Lorentz group to the gauge group which corresponds
to gravitational interaction. The PGTG leads to the change of gravitational
interaction in comparison with GR and Newton's theory of gravity at
cosmological scale, which are provoked by more complicated structure of
physical spacetime, namely by spacetime torsion \cite{a7,a8}.

Explicit form of gravitational equations of PGTG and their physical
consequences depend essentially on the structure of gravitational Lagrangian
${\cal L}_{\mathrm{g}}$, which is built by means of invariants of gravitational
gauge field strengths --- the curvature and torsion tensors. The most simple
PGTG is Einstein-Cartan theory based on the gravitational Lagrangian in the
form of scalar curvature of spacetime $U_4$ \cite{trautman}. In the frame of
Einstein-Cartan theory the DEP was discussed in \cite{capozz}, where some
phenomenological description of spinning matter was used. In connection with
this it should be noted that in the frame of Einstein-Cartan theory the torsion
is connected with spin momentum by linear algebraic relation and vanishes in
the case of spinless matter. Such situation seems unnatural by taking into
account that the torsion tensor is gravitational gauge strength corresponding
to transformations of translations, which are connected directly with
energy-momentum tensor in the frame of Noether formalism. The situation comes
to normal by including to ${\cal L}_{\mathrm{g}}$ terms quadratic in the
curvature and torsion tensors.

The most general form of the gravitational Lagrangian ${\cal L}_{\mathrm{g}}$
(without using Levi-Civita symbol) includes the linear in the scalar curvature
term as well as 9 quadratic terms (6 invariants of the curvature tensor and 3
invariants of the torsion tensor with indefinite parameters). The structure of
the gravitational equations and physical consequences of isotropic cosmology in
the frame of PGTG, in particular, the situation concerning the DEP depend
essentially on restrictions on indefinite parameters. The PGTG can be divided
into different sectors in dependence on the number of nonvanishing componets of
the torsion tensor and the order of the differential equations determining
their behaviour and behaviour of the scale factor that is connected with
restrictions on indefinite parameters of ${\cal L}_{\mathrm{g}}$.

In general case the torsion tensor for homogeneous isotropic models (HIM) is
described by two functions of time $S_1$ and $S_2$ determining trace and
pseudotrace of the torsion tensor respectively. At the first time the most
simple HIM with the only nonvanishing torsion function $S_1$ were built and
investigated in \cite{minkPL80}; it was shown that by certain restrictions on
equation of state of gravitating matter at extreme conditions (extremely high
energy densities and pressures) in the beginning of cosmological expansion all
cosmological solutions are regular with respect to metrics, Hubble parameter,
its time derivative and energy density by virtue of gravitational repulsion
effect at extreme conditions (see \cite{a11}). The regular Big Bang scenario
with inflationary stage in the beginning of cosmological expansion based on
such HIM was built and analyzed in \cite{a9,a10}.

Other sector of PGTG is so-called dynamical scalar torsion sector considered in
\cite{nester1,nester2}. HIM biult in this sector demonstrate the oscillating
behaviour of the Hubble parameter, and it is possible to obtain good
correspondence with SNe Ia observational data.

By investigation the sector of PGTG with two torsion functions it was shown
that the PGTG allows to explain the acceleration of cosmological expansion at
present epoch without using the notion of dark energy \cite{a12}.\footnote{At
the first time equations for HIM with two torsion functions were deduced in
\cite{a17}. These equations were considered in \cite{a18} with the purpose to
obtain their solutions; however, so called "modified double duality ansatz"
used in \cite{a18} by obtaining solutions with non-vanishing torsion function
$S_2$ is not applicable in this case even for the vacuum (see \cite{a19}) and
its application leads to incorrect solutions.}  This result is due to  the
fact, that the cosmological equations for HIM at asymptotics take  the form of
Friedmann cosmological equations of GR with effective cosmological constant
induced by spacetime torsion if certain restrictions on indefinite parameters
of gravitational Lagrangian are imposed. As it was shown in \cite{a13},
isotropic cosmology based on HIM with two torsion functions offers
opportunities to solve also the problem of dark matter. The analysis of regular
inflationary cosmological models built on the base of HIM with two torsion
functions by certain restrictions on indefinite parameters in gravitational
equations for such models was fulfilled in \cite{a14}. The present paper is
devoted to numerical analysis of HIM with two torsion functions of accelerating
Universe at asymptotics when energy densities are sufficiently small.

\section{\label{secii}Cosmological equations for homogeneous isotropic models}
In this Section we briefly repeat the derivation of the cosmological equations for HIM with two
torsion functions (see \cite{a12,a14}).

In the framework of PGTG the role of gravitational field variables play the tetrad $h^i{}_\mu$ and
the Lorentz connection $A^{ik}{}_\mu$; corresponding field strengths are the torsion tensor
$S^i{}_{\mu\nu}$ and the curvature tensor $F^{ik}{}_{\mu\nu}$ defined as
\[
S^i{}_{\mu \,\nu }  = \partial _{[\nu } \,h^i{}_{\mu ]}  - h_{k[\mu } A^{ik}{}_{\nu ]}\,,
\]
\[
F^{ik}{}_{\mu\nu }  = 2\partial _{[\mu } A^{ik}{}_{\nu ]}  + 2A^{il}{}_{[\mu } A^k{}_{|l\,|\nu
]}\,,
\]
where holonomic and anholonomic space-time coordinates are denoted by means of greek and latin
indices respectively.

We will consider the PGTG based on gravitational Lagrangian given in the following general form
\begin{eqnarray}\label{lagr}%\fl
{\cal L}_{\mathrm{g}}=  f_0\, F+F^{\alpha\beta\mu\nu}\left(f_1\:F_{\alpha\beta\mu\nu}+f_2\:
F_{\alpha\mu\beta\nu}+f_3\:F_{\mu\nu\alpha\beta}\right)+ F^{\mu\nu}\left(f_4\:F_{\mu\nu}+f_5\:
F_{\nu\mu}\right) \nonumber \\
+f_6\:F^2+S^{\alpha\mu\nu}\left(a_1\:S_{\alpha\mu\nu}+a_2\: S_{\nu\mu\alpha}\right)
+a_3\:S^\alpha{}_{\mu\alpha}S_\beta{}^{\mu\beta}, %\nonumber
\end{eqnarray}
where $F_{\mu\nu}=F^{\alpha}{}_{\mu\alpha\nu}$, $F=F^\mu{}_\mu$, $f_i$ ($i=1,2,\ldots,6$), $a_k$
($k=1,2,3$) are indefinite parameters, $f_0=(16\pi G)^{-1}$, $G$ is Newton's gravitational constant
(the velocity of light in the vacuum is equal to 1). Gravitational equations of PGTG obtained from
the action integral $ I = \int {\left({\cal L}_{\mathrm{g}} + {\cal L}_{\mathrm{m}} \right) \,}
h\,d^4 x$, where $h=\det{\left(h^i{}_\mu\right)}$ and ${\cal L}_{\mathrm{m}}$ is the Lagrangian of
gravitating matter, contain the system of 16+24 equations corresponding to gravitational variables
$h^i{}_\mu$ and $A^{ik}{}_\mu$. The sources of gravitational field in PGTG are the energy-momentum
and spin tensors.   In present paper we will consider perfect fluid with energy density $\rho$,
pressure $p=p(\rho)$ and vanishing spin tensor as a source of gravitational field.

High spatial symmetry of HIM allows to describe these models by three functions of time $t$: the
scale factor of Robertson-Walker metrics $R$ and two torsion functions $S_{1}$ and $S_{2}$
determining the curvature functions $A_{k}$ ($k=1,2,3,4$) as following
\begin{eqnarray}%\label{2}
%\eqalign{
&&
    A_1=\dot{H}+H^2-2HS_1-2\dot{S}_1,
        %\hfill
    \nonumber\\
&&    A_{2}  = \frac{k} {{R^2 }} + \left( {H - 2S_1 } \right)^2  - S_2^2,
        %\hfill
    \nonumber\\
&&    A_{3}  = 2\left( {H - 2S_1 } \right)S_2,
        %\hfill
    \nonumber\\
&&    A_{4}  = \dot S_2+HS_2,\nonumber %\hfill
%}
\end{eqnarray}
where $H=\dot{R}/R $ is the Hubble parameter and a dot denotes the differentiation with respect to
time. The system of gravitational equations of PGTG for HIM in considered case takes the following
form:
\begin{eqnarray}\label{1}%\fl
a\left( {H - S_1 } \right)S_1  - 2bS_2^2  - 2f_0 A_{2}  + 4f\left( {A_{1}^2 - A_{2}^2 } \right)
%\nonumber \\
+ 2q_2 \left( {A_{3}^2 -
A_{4}^2 } \right) =  - \frac{\rho } {3},\\
%\end{eqnarray}
%\begin{eqnarray}
\label{2}%\fl
a\left( {\dot S_1  + 2HS_1  - S_1^2 } \right) - 2bS_2^2  - 2f_0
\left( {2A_{1} + A_{2} } \right)
%\nonumber \\
- 4f\left( {A_{1}^2 - A_{2}^2 } \right) - 2q_2 \left( {A_{3}^2  - A_{4}^2 } \right) =
p,\\
%\end{eqnarray}
%\begin{eqnarray}
\label{3}%\fl
f\left[ {\dot A_{1}  + 2H\left( {A_{1}  - A_{2} } \right) + 4S_1 A_{2} } \right] + q_2 S_2 A_{3}
%\nonumber \\
- q_1 S_2 A_{4}  +
\left( {f_0  + \frac{a} {8}} \right)S_1  = 0,\\
%\end{eqnarray}
%\begin{eqnarray}
\label{4}%\fl
q_2 \left[ {\dot A_{4}  + 2H\left( {A_{4}  - A_{3} } \right) + 4S_1 A_{3} }
\right] - 4f\,S_2 A_{2}
%\nonumber \\
- 2q_1 S_2 A_{1}  - \left( {f_0  - b} \right)S_2  = 0,
\end{eqnarray}
where
\begin{eqnarray}
  a = 2a_1  + a_2  + 3a_3, \qquad b = a_2  - a_1,
%\hfill
\nonumber\\
  f = f_1  + \frac{{f_2 }} {2} + f_3  + f_4  + f_5  + 3f_{6}\, ,
%\hfill
\nonumber\\
  q_1  = f_2  - 2f_3  + f_4  + f_5  + 6f_{6}, \qquad q_2  = 2f_1  - f_2 .
%\hfill \\
\nonumber
\end{eqnarray}
The system of gravitational equations (\ref{1})--(\ref{4}) allows to obtain the cosmological
equations generalizing Friedmann cosmological equations of GR and equations for the torsion
functions $S_1$ and $S_2$.

To exclude higher derivatives of the scale factor $R$ from cosmological
equations the following restriction on indefinite parameters $a_k$ was imposed:
$a=0$ (see \cite{minkPL80,a19}). Obtained gravitational equations can be
simplified if an additional restriction on parameters $f_k$ is imposed
$2f=q_1+q_2$ \cite{a12}. Then cosmological equations and equations for torsion
functions contain three indefinite parameters: parameter
$\alpha=\frac{1}{3}\frac{f}{f_0^2}>0$ with inverse dimension of energy density,
parameter $b$ with dimension of parameter $f_0$ and dimensionless parameter
$\veps=\frac{q_2}{f}$. The particle content of the PGTG with these restrictions
on indefinite parameters of the gravitational Lagrangian (\ref{lagr}) was
discussed in ref.~\cite{a14}.

For further analysis, we transform cosmological equations to dimensionless form by introducing
dimensionless units for all variables and parameter $b$ entering these equations and denoted by
means of tilde:
\begin{equation}\label{dimless}
\begin{array}{lcl}
    t\to\tilde{t}=t/\sqrt{6 f_0 \alpha},& {}
            & R\to\tilde{R}=R/\sqrt{6f_0 \alpha},\\
    \rho\to\tilde{\rho}=\alpha\,\rho, & & p\to\tilde{p}=\alpha\,p,\\
    S_{1,2}\to\tilde{S}_{1,2}=S_{1,2}\sqrt{6f_0 \alpha}, & &
            b\to\tilde{b} = b/f_0, \\
    H\to\tilde{H}=H\sqrt{6f_0 \alpha}, & &
\end{array}
\end{equation}
where dimensionless Hubble parameter $\tilde{H}$ is defined  by usual way
$\tilde{H}=\tilde{R}^{-1}\frac{d \tilde{R}}{d \tilde{t}}$. As result cosmological equations
((22)--(23) in Ref.~\cite{a12}) take the following dimensionless form, where the differentiation
with respect to dimensionless time $\tilde{t}$ is denoted by means of the prime and the sign of \
$\tilde{}$\, is omitted below in this Section and Sections \ref{seciii} and \ref{nintsec}:
\begin{eqnarray}\label{gcfe1}
    \frac{k}{R^2}  +  (H-2S_1)^2 %\nonumber\\
    &=& \frac{1} {Z}
        \left[
            {\rho  +\left(Z- b\right) S_2^2
            + \frac{1}{4} \left( {\rho  - 3p - 2bS_2^2 } \right)^2 }
        \right]
\nonumber\\
        &&- \frac{\varepsilon}{2Z}
            \left[
                {\left( {HS_2  + S_2' } \right)^2
                + 4\left( {\frac{k}{{R^2 }} - S_2^2 } \right)S_2^2 }
            \right],
\end{eqnarray}
\begin{eqnarray}\label{gcfe2}
    H'  &+&  H^2 - 2HS_1 - 2S_1' %\nonumber \\
     =  -\frac{1} {2Z}
        \left[
            \rho  + 3p - \frac{1 } {2} \left( {\rho  - 3p - 2bS_2^2 } \right)^2
        \right]
\nonumber\\
        &  - & \frac{\varepsilon }{Z}\left( {\rho  - 3p - 2bS_2^2 } \right)S_2^2
%\nonumber\\
         + \frac{{\varepsilon }} {2Z}
            \left[ {\left( {HS_2  + S_2' } \right)^2
                + 4\left( {\frac{k}{{R^2 }} - S_2^2 } \right)S_2^2 }
            \right], %\\
\end{eqnarray}
\[
    \left(Z \equiv 1+\rho - 3p - 2\left( {b + \varepsilon } \right)S_2^2\right).\nonumber
\]
In considering case of HIM filled with usual gravitating matter with equation
of state in the form $p=p(\rho)$ the torsion function $S_{1}$ in dimensionless
form appearing in (\ref{gcfe1})--(\ref{gcfe2}) is
\begin{equation} \label{S1expr}
    S_1  =   -\frac{3}{4Z} \left\{
           H\left[\left(\rho+p\right)%
        \left(3\frac{dp}{d\rho}-1 \right) + 2\varepsilon S_2^2\right]
            -\frac{2}{3}\left( {2b - \veps } \right) S_2 \, S_2'
        \right\}
\end{equation}
and  dimensionless torsion function $S_{2}$ satisfies the following differential equation of the
second order:
\begin{eqnarray}\label{gcfe3}
    \varepsilon \left[ S_2''  + 3H S_2'  + 3H' S_2  - 4\left(S_1'  - 3 HS_1
        + 4S_1^2\right) S_2  \right]
\nonumber \\
        - 2\left( {\rho  - 3p - 2bS_2^2 } \right)S_2
        - 2\left( {1  - b}\right)S_2  = 0\,.
\end{eqnarray}
The conservation law for gravitating matter in dimensionless units has the
usual form
%\begin{equation}
\begin{equation}
\label{conslaw} \rho'+3H\left(\rho+p\right)=0.
\end{equation}
%\end{equation}

\section{\label{seciii}Critical points analysis}

The system of equations (\ref{gcfe2}) -- (\ref{gcfe3}) together with
conservation law (\ref{conslaw}) completely determine the dynamics of HIM, if
the equation of state of matter is given. For further analysis we will consider
flat model ($k=0$) filled with matter with barotropic equation of state
$p=w\rho$ $(w=const)$. The aforementioned system of equations can be
represented in the form of four first order differential equations for $H$,
$S_2$, $U=S_2'$ and $\rho$:
\begin{equation}\label{firstordersyst}
 M_0 \mathbf{Y}'=\mathbf{F},
\end{equation}
where the matrix $M_0$ is
\begin{equation}
 M_0=\left(
    \begin{array}{cccc}
     1-2\frac{\pat S_1}{\pat H} & -2\frac{\pat S_1}{\pat S_2} & -2\frac{\pat S_1}{\pat U} & -2\frac{\pat S_1}{\pat\rho}
        \\
     0 & 1 & 0 & 0
    \\
     3 \varepsilon S_2 - 4\varepsilon\frac{\pat S_1}{\pat H}S_2 & - 4\varepsilon\frac{\pat S_1}{\pat S_2}S_2
     & - 4\varepsilon\frac{\pat S_1}{\pat U}S_2
     & - 4\varepsilon\frac{\pat S_1}{\pat\rho}S_2
    \\
     0 & 0 & 0 & 1
    \end{array}
      \right)
\end{equation}
and
\[
 \mathbf{Y}=\left(
    \begin{array}{l}
     H\\
     S_2\\
     U\\
     \rho
    \end{array}
 \right), \qquad
 \mathbf{F}=\left(
    \begin{array}{l}
     \mathcal{F}_1(H,S_2,U,\rho)\\
     \mathcal{F}_2(H,S_2,U,\rho)\\
     \mathcal{F}_3(H,S_2,U,\rho)\\
     \mathcal{F}_4(H,S_2,U,\rho)
    \end{array}
 \right),
\]
\begin{eqnarray}
  \mathcal{F}_1 & = & -H^2 + 2HS_1
      -\frac{1} {2Z}
        \left\{
            \left(1+3w\right)\rho  - \frac{1 } {2} \left[ \left(1-3w\right)\rho  - 2bS_2^2 \right]^2
        \right\}
\nonumber\\
    & & -  \frac{\varepsilon }{Z}\left[ \left(1-3w\right)\rho  - 2\left(b-1\right)S_2^2  \right]S_2^2
%\nonumber\\
        %& &
        + \frac{{\varepsilon }} {2Z}
            \left( {HS_2  + U } \right)^2,\\
  \mathcal{F}_2 & = & U,\\
  \mathcal{F}_3 & = & - \varepsilon \left[3H U  + 4\left(3 HS_1
        - 4S_1^2\right) S_2  \right]
        + 2\left[\left(1-3w\right) \rho  - 2bS_2^2  \right]S_2
%\nonumber \\
    %& &
        + 2\left( {1  - b}\right)S_2 ,\\
  \mathcal{F}_4 & = & -3 \left(1+w\right)\rho H.
\end{eqnarray}
The function $S_1$ can be written as
\begin{equation} \label{S1expr2}
    S_1  =   -\frac{3}{4Z} \left\{
           H\left[\left(1+w\right)%
        \left(3w-1 \right)\rho + 2\varepsilon S_2^2\right]
            -\frac{2}{3}\left( {2b - \veps } \right) S_2 U
        \right\}.
\end{equation}

Critical points $P_i=P_i(H_\mathrm{c}, S_{2\mathrm{c}}, U_\mathrm{c}, \rho_\mathrm{c})$ of the
first order system of differential equations (\ref{firstordersyst}) can be obtained by setting
$H'$, $S_2'$, $S_1'$, $\rho'$ to zero \cite{agarwal,arnold}, i.e. by solving the following system
of equations:
\begin{equation}\label{specpoint}
 \mathcal{F}_i(H, S_2, U, \rho)=0, \qquad \left(i=1,\ldots,4\right).
\end{equation}
In the case of flat model ($k=0$), solutions of (\ref{specpoint}) have to satisfy (\ref{gcfe1})
with $k=0$. Eq. (3.4) leads to $U_\mathrm{c}=0$.

Obviously, the point  $P_0$ with vanishing values of $H_\mathrm{c}, S_{2\mathrm{c}},
\rho_\mathrm{c}$ satisfies (\ref{specpoint}). Analogously to GR this point is the point of
complicated equilibrium. To analyze the stability of other critical points
$P(H_\mathrm{c},S_{2\mathrm{c}},0,\rho_\mathrm{c})$ satisfying (\ref{specpoint}) it is necessary to
build linearized form of the system (\ref{firstordersyst}). Near the critical point the variables
can be written in the form $H=H_\mathrm{c}+\Delta H$, $S_2=S_{2\mathrm{c}}+\Delta S_2$, $U=\Delta
U$, $\rho=\rho_\mathrm{c}+\Delta\rho$ and the linearization of the system (\ref{firstordersyst})
takes the following relation
\begin{equation}\label{firstordersystapprox}
 \Delta\mathbf{Y}'=M_0^{-1}M_1\,\Delta\mathbf{Y},
\end{equation}
where the components of the matrix $M_1$ are given by
\[
 M_{1,ij}=\left.\left(\frac{\pat \mathcal{F}_i}{\pat Y_j}\right)\right|_{P}.
\]
Stability of the point $P$ is determined by the eigenvalues $\lambda_i$ of the matrix $M_0^{-1}M_1$
\cite{agarwal,arnold}. Characteristic equation $\det\left(M_1-\lambda M_0\right)=0$ leads to
quartic expression with respect to $\lambda$, which can be written as
\begin{equation}\label{charactereq}
  \lambda^4+c_1\lambda^3+c_2\lambda^2+c_3\lambda+c_4=0.
\end{equation}
If the real parts of all $\lambda_i$ is negative, then the critical point $P$ is stable and the
gravitational equations (\ref{gcfe2}) -- (\ref{conslaw}) can have asymptotics to this point $H\to
H_\mathrm{c}$, $S_2\to S_{2\mathrm{c}}$, $S'_2 \to 0$, $\rho \to \rho_\mathrm{c}$ at $t\to
+\infty$.

According to the Routh-Hurwitz theorem  all $\lambda_i$ will have negative real parts if the main
minors of the matrix
\begin{equation}\label{RHmatrix}
 \left(
  \begin{array}{cccc}
   c_1 & 1 & 0 & 0\\
   c_3 & c_2 & c_1 & 1\\
   0 & c_4 & c_3 & c_2 \\
   0 & 0 & 0 & c_4
  \end{array}
 \right)
\end{equation}
are positive \cite{agarwal}, i.e.
\begin{equation}\label{stabilityineq}
 c_1>0,\qquad c_1c_2-c_3>0,\qquad c_1c_2c_3-c_1^2c_4-c_3^2>0\qquad \text{and}\qquad c_4>0.
\end{equation}

The equation $\mathcal{F}_4(H_\mathrm{c},S_{2\mathrm{c}}, 0,\rho_\mathrm{c})=0$
gives two kinds of critical points: with vanishing Hubble parameter
$H_\mathrm{c}=0$ or with vanishing energy density $\rho_\mathrm{c}=0$. Let us
consider in details both of them.

\subsection{Critical points with vanishing Hubble parameter}

If $H_\mathrm{c}=0$ the system (\ref{specpoint}) is reduced to the system of two algebraic
equations
\begin{eqnarray}\label{specpoint1a}
  & & \frac{1} {2}
        \left\{
            \left(1+3w\right)\rho  - \frac{1 } {2} \left[ \left(1-3w\right)\rho  - 2bS_2^2 \right]^2
        \right\}
      +  \varepsilon\left[ \left(1-3w\right)\rho  - 2\left(b-1\right)S_2^2  \right]S_2^2 = 0,\\
  & &   \left[1  - b +\left(1-3w\right) \rho  - 2bS_2^2 \right]S_2 = 0.
  \label{specpoint1b}
\end{eqnarray}
Generally speaking, except trivial solution $S_{2\mathrm{c}}=0$, $\rho_\mathrm{c}=0$ this system
admits non-zero solutions for $S_{2\mathrm{c}}$ and $\rho_\mathrm{c}$. According to (3.14) there
are the following possibilities: $S_2=0$ and $S_2\neq 0$. In the first case we obtain the trivial
solution with $\rho_c=0$. If $S_{2\mathrm{c}}\neq 0$, from (\ref{specpoint1b}) it follows that
$(1-3w)\rho=b-1+2bS_2^2$. As result we have
\begin{equation}\label{specpoint1result}
 S_1=0\qquad \text{and}\qquad Z=b-2\varepsilon S_2^2.
\end{equation}
Substitution (\ref{specpoint1result}) into (\ref{gcfe1}) gives
\begin{equation}
 \frac{k}{R^2}\left(1+\frac{2\varepsilon}{Z}\right)-\frac{1}{Z}\left[\rho+\frac{1}{4}\left(b-1\right)^2\right]=0,
\end{equation}
and this equation does not have non-trivial solution in physical space ($\rho>0$) for flat models
($k=0$).

\subsection{Critical points with vanishing energy density}

If $\rho_\mathrm{c}=0$ the system (\ref{specpoint}) is reduced to the system of two algebraic
equations
\begin{eqnarray}
  & & -H^2 + 2HS_1
      +\frac{b^2} {Z}  S_2^4  +  \frac{2\varepsilon\left(b-1\right)}{Z}S_2^4
      + \frac{{\varepsilon }} {2Z} H^2S_2^2=0,
 \label{specpoint2a}\\
  & &    \left[2\varepsilon \left(3 HS_1 - 4S_1^2\right) + 2 b S_2^2  - \left(1  - b\right)\right]S_2 = 0,
  \label{specpoint2b}
\end{eqnarray}
where the functions $S_1$ and $Z$ can be represented in the following form
\begin{equation} \label{S1expr2st1}
    S_1  =   -\frac{3\varepsilon}{2Z} H S_2^2
    \qquad \text{and} \qquad
    Z = 1 - 2\left( {b + \varepsilon } \right)S_2^2.
\end{equation}
Neglecting the case $S_2=0$, it is possible to obtain from (\ref{specpoint2a}) --
(\ref{specpoint2b}) equation for $H$ in closed form. To do this, the system of equations
(\ref{specpoint2a})--(\ref{specpoint2b}) can be rewritten in the following form
\begin{eqnarray}
  & & 2\left(1-2bS_2^2\right)H^2+\varepsilon H^2S_2^2-2\left[b^2+2\varepsilon\left(b-1\right)\right]S_2^4=0,
 \label{specpoint2a1}\\
  & & 9\varepsilon\left[1-2\left(b+\varepsilon-\varepsilon^2\right)\right]H^2S_2^2-2bS_2^2Z^2+\left(1-b\right)Z^2=0.
  \label{specpoint2b1}
\end{eqnarray}
Polynomials in the left-hand-side of equations (\ref{specpoint2a1})--(\ref{specpoint2b1}) generates
ideal in the polynomial ring in the variables $H$ and $S_2$ \cite{cox}. There are different ways to
choose the basis in the ideal of polynomials. One of them is a Gr\"obner basis with lexicographic
ordering $H\prec S_2$. Using computer algebra system \textit{Wolfram Mathematica} the first element
of Gr\"obner basis of aforementioned polynomial ring takes the form
\begin{eqnarray}\label{criticalH}
 & &
 \left(-9 \varepsilon  (b+2 \varepsilon ) H^2 +b^2+2 b \varepsilon -2 \varepsilon \right)
 \left\{ 2 b \varepsilon ^2 \left[8 b^2-b \varepsilon  (9 \varepsilon +2)+8 \varepsilon ^2\right] H^6
 \right.\nonumber \\
 & &
 -\varepsilon  \left[4 b^4 (\varepsilon -2)+b^3 \varepsilon  (17 \varepsilon +22)
 -b^2 \varepsilon  \left(32 \varepsilon ^2+41 \varepsilon +34\right)+8 b \varepsilon ^2 (8 \varepsilon +3)
 -32 \varepsilon ^3\right]  H^4
 \nonumber \\
 & &
 -2  \left[b^5 (\varepsilon +4)+b^4 (5-6 \varepsilon ) \varepsilon
 -2 b^3 \varepsilon  \left(8 \varepsilon ^2-4 \varepsilon +3\right)+2 b^2 \varepsilon ^2 (24 \varepsilon +1)
 \right. \nonumber \\
 & & \left.\left.
 -4 b \varepsilon ^2 (12 \varepsilon +1)+16 \varepsilon ^3\right] H^2
 +2 (b-1)^2 \left(b^2+2 b \varepsilon -2 \varepsilon \right)^2 \right\} = 0.
\end{eqnarray}
Repeating this procedure with lexicographic ordering $S_2\prec H$ we have the first element of
Gr\"obner including only $S_2$
\begin{eqnarray}
 \label{criticalS2}
 \left[2 (b+2 \varepsilon ) S_2^2-1\right] \left\{ 2 b \left[8 b^2-b \varepsilon  (9 \varepsilon +2)+8 \varepsilon ^2\right] S_2^6
 +\left[8 b^3-2 b^2 (\varepsilon +12)
 \right.\right. \nonumber \\
 \left.\left.
    +4 b \varepsilon  (2 \varepsilon +1)-8 \varepsilon ^2 \right] S_2^4
 -\left[8 b^2+b (\varepsilon +12)+\varepsilon \right] S_2^2 +2 (b-1) \right\} &=& 0.
\end{eqnarray}
According to general mathematical theorems \cite{cox} roots of the system of equations
(\ref{specpoint2a1})--(\ref{specpoint2b1}) turn  (\ref{criticalH}) and (\ref{criticalS2}) into
true.

Analytic analysis of stable points determined by the system
(\ref{specpoint2a})--(\ref{specpoint2b}) is possible approximately only if $1-b\to +0$ and
$\varepsilon\to 0$. In other cases it is necessary to use numerical methods.

\subsubsection[Approximate analysis in the case $0<1-b<<1$]{Approximate analysis in the case $0<1-b\ll 1$}

System of equations (\ref{specpoint2a}) -- (\ref{specpoint2b}) admits simple approximate solution
if $0<1-b\ll 1$. This solution was initially obtained in ref.~\cite{a12} and after transformation
 to dimensionless form (\ref{dimless}) it reads:
\begin{equation}
H_\mathrm{c}= \frac{1-b}{2\sqrt{b}}, \qquad S_{2\mathrm{c}}=\sqrt{\frac{1-b}{2b}}.
\end{equation}
This approximation is valid up to cubic term in $1-b$ and linear term in $\varepsilon$.

In the case  $0<1-b\ll 1$ the stability of the critical point
$P_2\approx\left(\frac{1-b}{2\sqrt{b}}, \sqrt{\frac{1-b}{2b}},0,0\right)$ can be analyzed
analytically. In this case the inequalities (\ref{stabilityineq}) leads to:
\begin{equation}
 \varepsilon>0, \quad w>-1.
\end{equation}

\subsubsection{Numerical analysis of stability}

\begin{figure}[t]
\begin{minipage}{0.48\textwidth}\centering{
 \includegraphics[width=\linewidth]{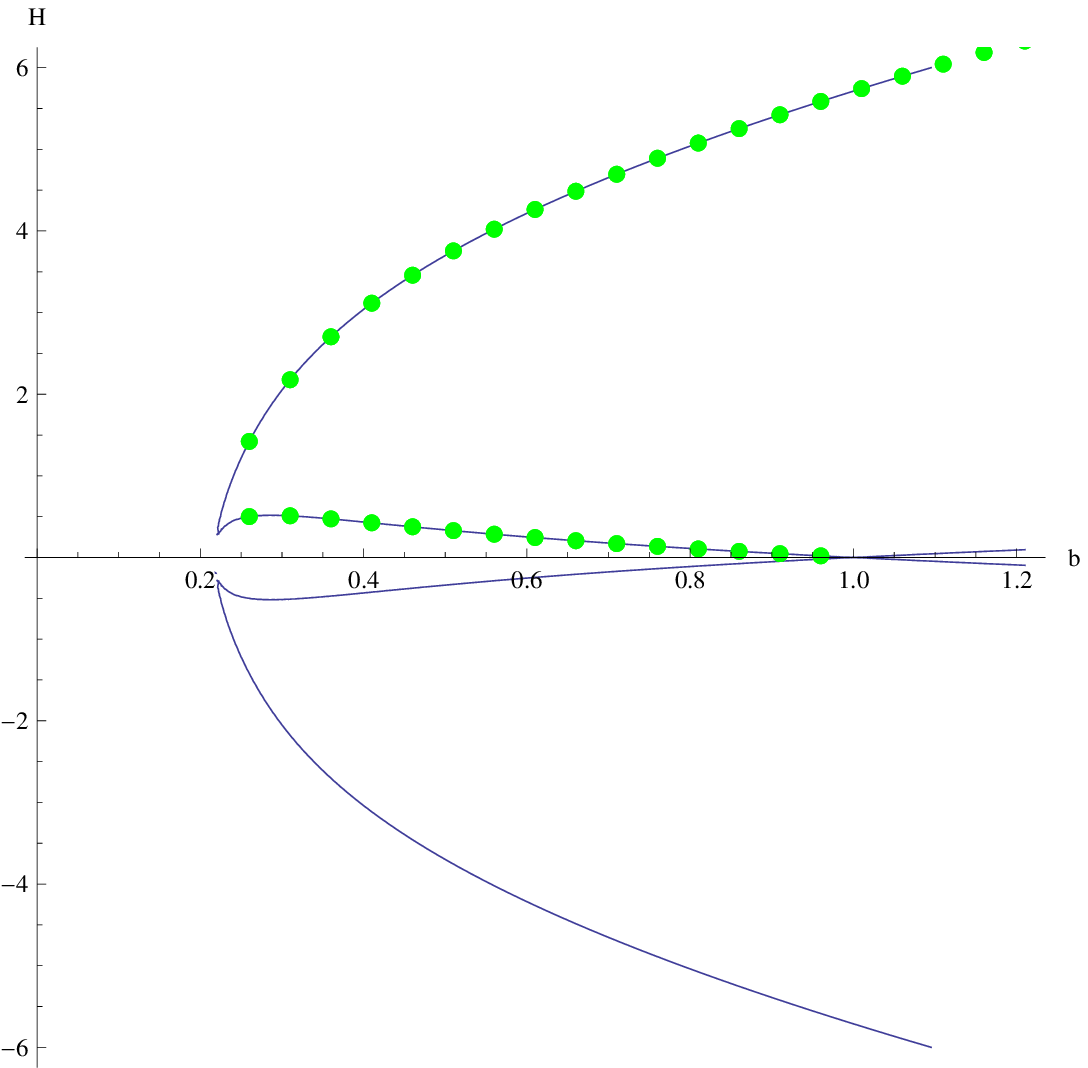}
}
\end{minipage}\, \hfill\,
\begin{minipage}{0.48\textwidth}\centering{
 \includegraphics[width=\linewidth]{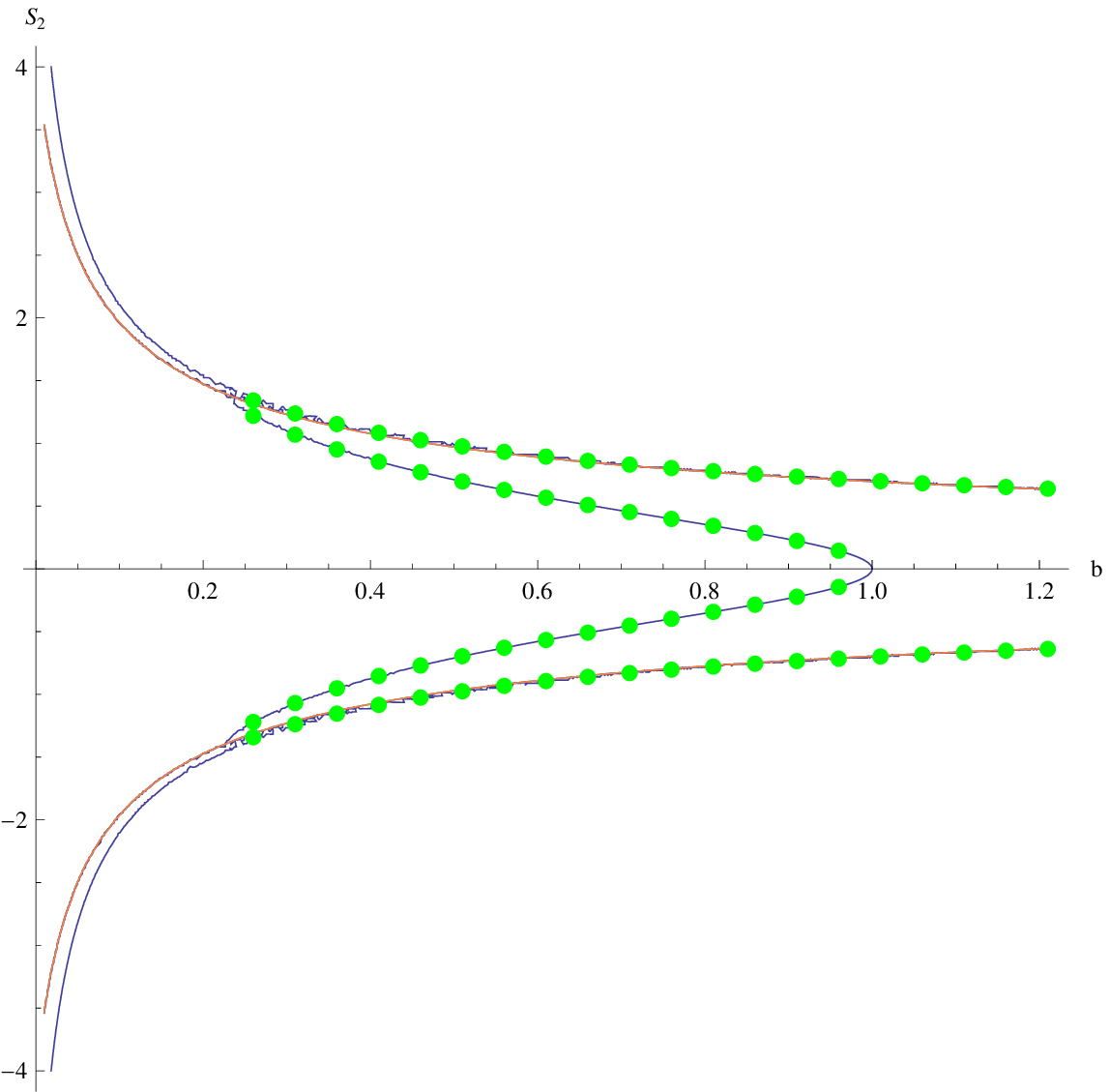}
}
\end{minipage}
\caption{\label{figstable1}$H_\mathrm{c}$ and $S_{2\mathrm{c}}$ as function of $b$
($\varepsilon=0.03$). Stable critical points are marked by (green) circle. Unstable critical points
and points of complicated equilibrium are not shown in  this figure. Solid lines in the left and
right panel of figure correspond to solutions of eqs. (\ref{criticalH}) and (\ref{criticalS2})
respectively.}
\end{figure}

As an exact analytic expression for solution of the system (\ref{specpoint2a})--(\ref{specpoint2b})
does not exist in general case, it is necessary to use numerical methods to analyze stability of
the critical points. The procedure of the numerical analysis of the stability points is following.
\begin{enumerate}
 \item For given value of $\varepsilon$, the system (\ref{specpoint2a})--(\ref{specpoint2b})
 is solved numerically for the set of values $b$.
 \item For every real solution of the system (\ref{specpoint2a})--(\ref{specpoint2b}) at given values
 of $\varepsilon$ and $b$ characteristic equation $\det\left(M_1-\lambda M_0\right)=0$ has to be solved
 with respect to $\lambda$.
 \item The real parts of obtained $\lambda_i$ have to be tested for negativity.
\end{enumerate}
For example, the results of this procedure for $\varepsilon=0.03$ are given in
figure~\ref{figstable1}. The calculation are performed for $b$ varying from $0{.}01$ to $1{.}2$
with a step $\Delta b=0.05$. In the left panel of figure~\ref{figstable1} the curves determined by
(\ref{criticalH}) are imposed. In the right panel an analogous curves for $S_2$ determined by
(\ref{criticalS2}) are imposed.

From figure~\ref{figstable1} it is possible to see, that there is minimal value of $b$ assuming
nontrivial solution of eqs. (\ref{specpoint2a})--(\ref{specpoint2b}). This value can be found by
setting $H$ to zero in (\ref{criticalH}). As result we have the following restriction on $b$
\begin{equation}
 b>-\varepsilon+\sqrt{\varepsilon\left(2+\varepsilon\right)}.
\end{equation}

\section{Numerical integration of the system of gravitational equations}\label{nintsec}

In this paper we will analyze the late time behaviour of the solution of the system
(\ref{gcfe2})--(\ref{conslaw}). To make comparison with GR we will perform numerical integration of
the system of the gravitational equations for dust matter ($w=0$). To simulate late-time behaviour
the initial conditions  will be taken at $t_0=0$ according to the following procedure.
\begin{enumerate}
 \item For given value of $\varepsilon$ and $b$
 ($\varepsilon>0$, $-\varepsilon+\sqrt{\varepsilon\left(2+\varepsilon\right)}<b<1$)
 algebraic system (\ref{specpoint2a})--(\ref{specpoint2b}) is solved numerically and all critical points
 $P_i(H_\mathrm{c}, S_{2\mathrm{c}},0,0)$ are found.  Only real solutions  are considered.
 \item For every critical point, the stability analysis is carried out according to the previous subsection and stable
 point with minimal  positive $H_\mathrm{c}$ and positive $S_{2\mathrm{c}}$ is selected.\footnote{These
 values of $H_\mathrm{c}$ and  $S_{2\mathrm{c}}$ correspond to the vacuum as de Sitter spacetime
 with torsion \cite{a19}.}
 \item The torsion function $S_2$ and the Hubble parameter $H$ can be represented in the form
 \begin{eqnarray}
  \label{Hsubst1}\label{Happrox}
  & & H^2=H_\mathrm{c}^2+y_1\,\rho,\\
  \label{S2subst2}
  & & S_2^2=S_{2\mathrm{c}}^2+y_2\,\rho,
 \end{eqnarray}
 with some coefficients $y_1$ and $y_2$.\footnote{Further in this paper we will refer representation (\ref{Hsubst1})--(\ref{S2subst2})
 of $H$ and $S_2$  as late-time approximation.}  As the stable point is selected, then $\rho$ tends to zero at $t\to +\infty$.
 Keeping linear terms in $\rho$  the conservation law (\ref{conslaw}) can written as
 \begin{equation}\label{conslawapprox}
  \rho'=-3H_{\mathrm{c}}\rho.
 \end{equation}
 Substitution of (\ref{Hsubst1})--(\ref{conslawapprox}) into (\ref{gcfe2})--(\ref{gcfe3}) together with keeping terms linear in $\rho$
 gives two algebraic equations for determination of $y_1$ and $y_2$. Numerical solution of these algebraic
 equations for given  $\varepsilon$, $b$, $H_\mathrm{c}$ and $S_{2\mathrm{c}}$ gives $y_1$ and
 $y_2$.
 \item Positivity of obtained values of $y_1$ and $y_2$ is considered as applicability of the late time approximation (\ref{Hsubst1})--(\ref{S2subst2})
 and successful choice of stable critical point made in step 2 of current procedure. Further steps are performed only if $y_1>0$ and
 $y_2>0$.
 \item Initial condition for $\rho_0=\rho(t_0)$ is taken from the following equation
 \begin{equation}
  \frac{H^2(t_0)}{H_\mathrm{c}^2}\equiv\frac{H_\mathrm{c}^2+y_1\rho_0}{H_\mathrm{c}^2}=\frac{1}{\Omega_\Lambda},
 \end{equation}
 as result we have $H_0=H(t_0)=\sqrt{H_{\mathrm{c}}^2+y_1\,\rho_0}$ and $S_{20}=S_2(t_0)=\sqrt{S_{2\mathrm{c}}^2+y_2\,\rho_0}$.
 Here $\Omega_\Lambda$ is an additional free parameter that specifies initial conditions.
 \item Initial condition for $S_{20}'=S_2'(t_0)$ is obtained from (\ref{gcfe1}) taking into account $k=0$.
 The minimal in modulus value   of $S_{20}'$ is taken as initial value.
 \item For this choice of the parameters $\varepsilon$, $b$ and initial conditions $\rho_0$, $H_0$, $S_{20}$ and $S_{20}'$ the system
 of differential equations (\ref{gcfe2})--(\ref{conslaw}) is integrated numerically.
\end{enumerate}

\begin{figure}[t]
\begin{minipage}{0.48\textwidth}\centering{
 \includegraphics[width=\linewidth]{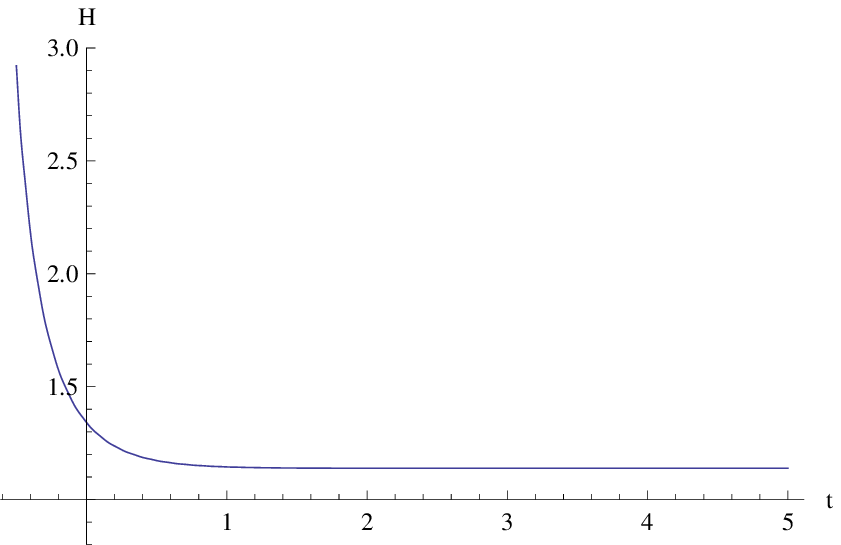}
}
\end{minipage}\, \hfill\,
\begin{minipage}{0.48\textwidth}\centering{
 \includegraphics[width=\linewidth]{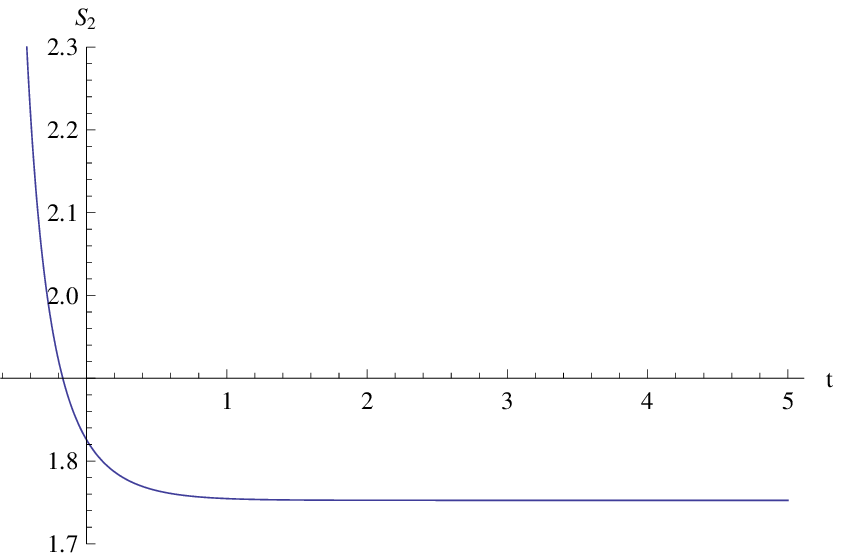}
}
\end{minipage}
\caption{\label{figm1}Late-time behaviour of Hubble parameter and $S_2$ torsion function.}
\end{figure}

\begin{figure}[t]
\begin{minipage}{0.48\textwidth}\centering{
 \includegraphics[width=\linewidth]{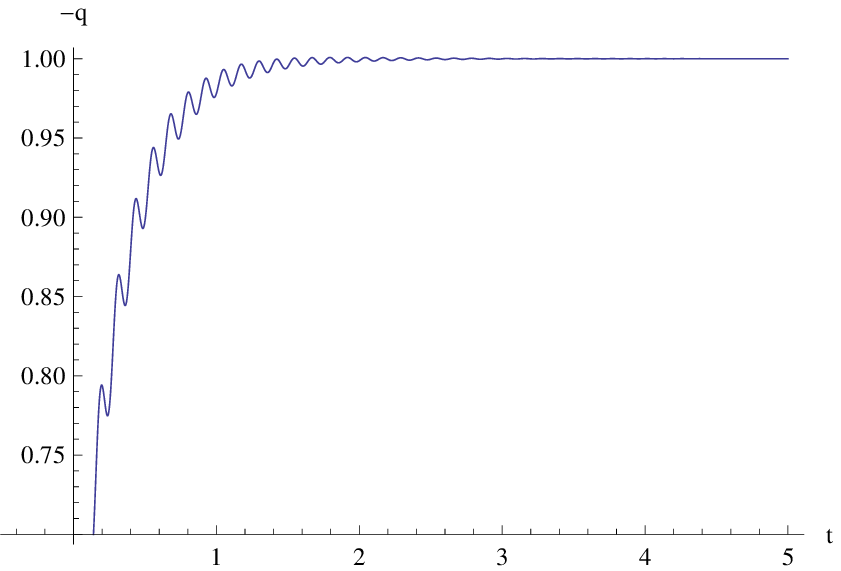}
}
\end{minipage}\, \hfill\,
\begin{minipage}{0.48\textwidth}\centering{
 \includegraphics[width=\linewidth]{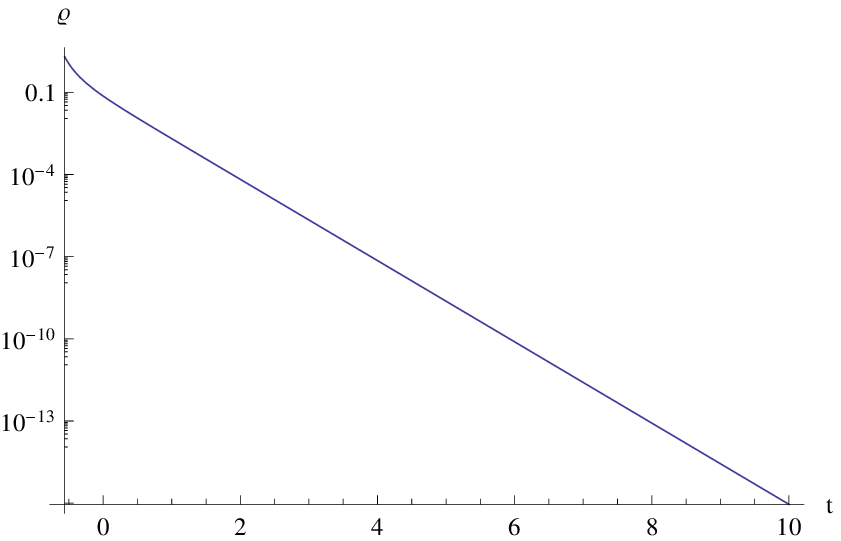}
}
\end{minipage}
\caption{\label{figm2}Late-time behaviour of the deceleration parameter and energy-density.}
\end{figure}

As an example let us consider the numerical solution at the following parameters and initial
conditions $\veps=0.00019$, $b=0.14$, $H_0=1.3417$, $S_{20}=1.82574$, $S'_{20}=-0.288424$,
$\rho_0=0.0730922$. This choice of the initial conditions gives
$H^2(t_0)/H^2(\infty)=1/\Omega_\Lambda=1/0{.}72$. Figures~2--3 show the behaviour of Hubble
parameter $H$, torsion function $S_2$, deceleration parameter
\begin{equation}\label{descparam}
 q=-R''R/{R'}^2
\end{equation}
and energy-density of dust matter $\rho$. Another choice of the initial conditions leads generally
speaking to cosmological solutions with another behaviour of the Hubble parameter and torsion
function $S_2$ because of their oscillating character.

\section{Comparison with observational data}

In this Section we will analyze what restrictions on indefinite parameters leads to solutions
corresponding to observational  data. It should be noted that the numerical solution of
gravitational equations allows to obtain time dependence of Hubble parameter $H(t)$ and scale
factor $R(t)$ for given values of parameters of gravitational Lagrangian and initial conditions. As
result dimensionless luminosity distance $\tilde{d}_{\mathrm{L}}$ can be obtained as a function of
redshift $z$
\begin{equation}\label{redshift}
 z=\frac{R(t_0)}{R(t)}-1=\frac{\tilde{R}(\tilde{t}_0)}{\tilde{R}(\tilde{t})}-1
\end{equation}
in the following form \cite{weinberg, riess}:
\begin{equation}\label{lumdist1}
 \tilde{d}_{\mathrm{L}}=\tilde{R}(\tilde{t}_0)(1+z)\int_{\tilde{t}}^{\tilde{t}_0}\frac{d\tilde{t}}{\tilde{R}(\tilde{t})}
    =(1+z)\int_0^{z}\frac{dz}{\tilde{H}(z)}.
\end{equation}

The predicted distance modulus $\mu=m-M$ ($m$ and $M$ are apparent and absolute magnitude
respectively) can be written as function of dimensional luminosity distance
$d_{\mathrm{L}}=\tilde{d}_{\mathrm{L}}\sqrt{6f_0\alpha}$ in the following form
\begin{equation}\label{distmodulus}
 \mu\equiv m-M=25+5\log_{10} d_{\mathrm{L}},
\end{equation}
where $d_\mathrm{L}$ is given in megaparsecs.

\subsection{Matching the late time approximation}
At first we will compare late time approximation (\ref{Happrox}) of $H(t)$ with supernovae type Ia
(SNe IA) observation data and predictions of the standard Big Bang Nucleosynthesis theory (BBN).
Dependence of the energy density of the dust matter $\rho(t)\sim R^3(t)$ allows to write Hubble
parameter in the late time approximation (\ref{Happrox}) as function of the redshift $z$. As result
the corresponding expression for the predicted distance modulus $\mu$ takes the form
\begin{equation} \label{distmodulus1}
 \mu(z) = 25+5\log_{10}\left[\left(1+z\right)\int_0^{z}\frac{dz}{\sqrt{\Omega_\Lambda+\left(1-\Omega_\Lambda\right)\left(1+z\right)^3}}\right]
    +\mu_0,
\end{equation}
where
\[
 \mu_0=5\log_{10}\frac{\sqrt{6f_0\alpha}}{\tilde{H}_0\cdot\mathrm{1Mpc}}.
\]

\subsubsection{Comparison with SNe Ia observational data}

We will start from comparison with supernovae observations using the Union2 compilation of 557 SNe
Ia data \cite{snIaUnion2} (see also \cite{snIaUnion}) and minimize
\begin{equation}\label{chi1}
 \chi_{\mathrm{SN}}^2=\sum_{i=1}^{557}\frac{\left[\mu(z_i)-\mu_{\mathrm{obs}}(z_i)\right]^2}{\sigma_i^2},
\end{equation}
where $\sigma_i$ is the distance modulus errors.

Best fit parameters $\Omega_\Lambda=0.73$ and $\mu_0=18{.}16$ gives $\chi_{\mathrm{SN}}^2=542.683$
and
 $\chi_{\mathrm{SN}}^2/\mathrm{dof}=0.974$ (dof --- degree of freedom). The value of $\mu_0=18{.}16$ corresponds to
 $70\,\mathrm{km}/(\mathrm{sec}\cdot\mathrm{Mpc})$ for Hubble constant at present epoch as
 in $\Lambda\mathrm{CDM}$-model of GR. Indeed, late time approximation (\ref{Happrox}) of $H(t)$ is similar to Friedmann equation of GR,
but differs only by effective gravitational constant $G_\mathrm{eff}$ determined as
$G_\mathrm{eff}=y_1 G$.

If the matter content includes baryonic and dark matter with relative  contributions
$\Omega_\mathrm{B}$ and  $\Omega_\mathrm{DM}$ to the total energy density, than it is easy to show
\cite{a13} that
 \begin{equation}\label{y1}
  y_1=\frac{1-\Omega_\mathrm{\Lambda}}{\Omega_\mathrm{DM}+\Omega_\mathrm{B}}.
 \end{equation}
In particular, as matter candidate for cold dark matter is not found yet, it is possible to fit SNe
Ia observational data in the discussed model  without using dark matter ($\Omega_\mathrm{DM}=0$).

 \subsubsection{Comparison with SNe Ia + BBN data}

 Calculations in the frame of standard BBN theory predicts
 $\omega_{\mathrm{B}}\equiv\Omega_\mathrm{B}h^2=0.0212 \pm 0.0010$ for
 baryon mass density \cite{steigman}, where  $h$ is the Hubble constant at present epoch in units
 of $100\,\mathrm{km}/(\mathrm{sec}\,\mathrm{Mpc})$. For $h=0{.}7$ obtained earlier the corresponding
 $\Omega_\mathrm{B}=0{.}043$. Assuming that the rate of light element production during Big Bang
 Nucleosynthesis does not depend on the presence of the torsion\footnote{This assumption does not
 contradict to results obtained in \cite{bruggen}.} and the dynamics of $H(t)$ from BBN epoch to present
 epoch is well approximated by (\ref{Happrox}), equation (\ref{y1}) gives $y_1=6{.}3$ for
 $\Omega_\mathrm{\Lambda}=0{.}73$ and $\Omega_\mathrm{DM}=0$. As $y_1$ is completely  determined
 by parameters $b$ and $\veps$, it seems  impossible to determine $b$ and $\veps$ simultaneously
 using late time approximation (\ref{Happrox}).

\subsection{Matching general case}

Besides approximation (\ref{Happrox}) there is another way to obtain the best fit parameters of
considered theory. Namely, for given values of $\varepsilon$, $b$ and specified initial conditions,
the procedure of numerical integration of exact system of differential equations
(\ref{gcfe2})--(\ref{conslaw}) allows to obtain solution for $H(t)$, $\rho(t)$ and predicted
distance modulus as a function of redshift $\mu=\mu(z)$. Obtained distance modulus - redshift
dependence $\mu(z)$ allows to calculate joined $\chi^2$ for Union2 data set and BBN predictions
\begin{equation} \label{chiall}
 \chi^2 =  \chi_{\mathrm{SN}}^2+\chi_{\mathrm{BBN}}^2,
\end{equation}
where
\[
 \chi_{\mathrm{BBN}}^2=\frac{\left(\Omega_\mathrm{B} 10^{-\bar{\mu}_0/5}c/\left(100\,\mathrm{km/sec}\right)-\omega_{\mathrm{B}}\right)^2}%
    {\sigma_{y_{\mathrm{B}}}^2},
\]
$c$ is the velocity of light and for computational purposes the functions
$\mu(z)$ and $\Omega_\mathrm{B}$  are written in the following form
\begin{equation} \label{distmodulus2}
 \mu(z) = 25+5\log_{10}\left[\left(1+z\right)\int_0^{z}\frac{dz}{\tilde{H}(z)}\right]
    +\bar{\mu}_0,
\end{equation}
\[
 \bar{\mu}_0=5\log_{10}\frac{\sqrt{6f_0\alpha}}{\mathrm{1Mpc}},
\]
\[
 \Omega_\mathrm{B}=\frac{1-\Omega_\mathrm{\Lambda}}{y_1}-\Omega_\mathrm{DM}.
\]
According to \cite{steigman} we will use $\omega_{\mathrm{B}}=0.0212$ and
$\sigma_{y_{\mathrm{B}}}=0.0010$.

In general case $\chi^2$  depends on parameters $\alpha$, $b$, $\veps$, initial conditions
$\rho_0$, $H_0$, $S_{20}$, $S'_{20}$ ($\Omega_\Lambda$) and cold dark matter contribution
$\Omega_\mathrm{DM}$ to the total energy density. The task of minimization of total $\chi^2$
implies the minimization with respect to all parameters and initial conditions. To simplify this
problem we will restrict the task by setting initial conditions in dependence on the values of
parameters as was discussed in Section~\ref{nintsec}.

Following the previous subsection at first we will use the Union2 compilation set only. Considering
the grid in the parameter region $0.00009\le \varepsilon\le 0.0002$, $0.12\le\tilde{b}\le 0.2$,
$0.66\le \Omega_{\Lambda}\le 0.74$ ($\Delta\varepsilon=10^{-5}$, $\Delta\tilde{b}=0.01$,
$\Delta\Omega_{\Lambda}=0.01$) and minimizing $\chi_{\mathrm{SN}}^2$ we find the minimum of
$\chi_{\mathrm{SN}}^2$ at $\varepsilon \approx 0.00019$ and $\tilde{b}\approx 0.12$,
$\Omega_{\Lambda}\approx 0.72$ and $\bar{\mu}_0=19.0074$ ($\chi_{\mathrm{SN}}^2 = 542.8$,
$\chi_{\mathrm{SN}}^2/\mathrm{dof}=0.975$). Obtained values of parameters correspond to
$70.0\,\mathrm{km}/(\mathrm{sec}\cdot\mathrm{Mpc})$ for Hubble constant at present epoch,
$\omega_\mathrm{B}=0.0173$ and $\alpha^{-1}=4.22\cdot 10^{-30}\,\mathrm{g}/\mathrm{cm}^3$. Obtained
value of $\omega_\mathrm{B}$ seems to be small and contradicts data on D- and ${}^3$He-abundance,
but close to data on ${}^4$He \cite{steigman2}.

\begin{figure}[t]
\begin{minipage}{0.48\textwidth}\centering{
 \includegraphics[width=\linewidth]{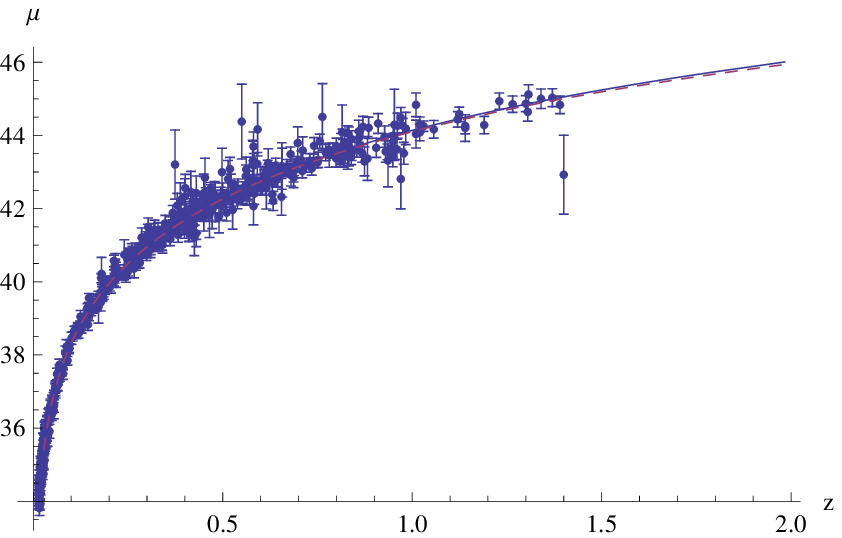}
}
\end{minipage}\, \hfill\,
\begin{minipage}{0.48\textwidth}\centering{
 \includegraphics[width=\linewidth]{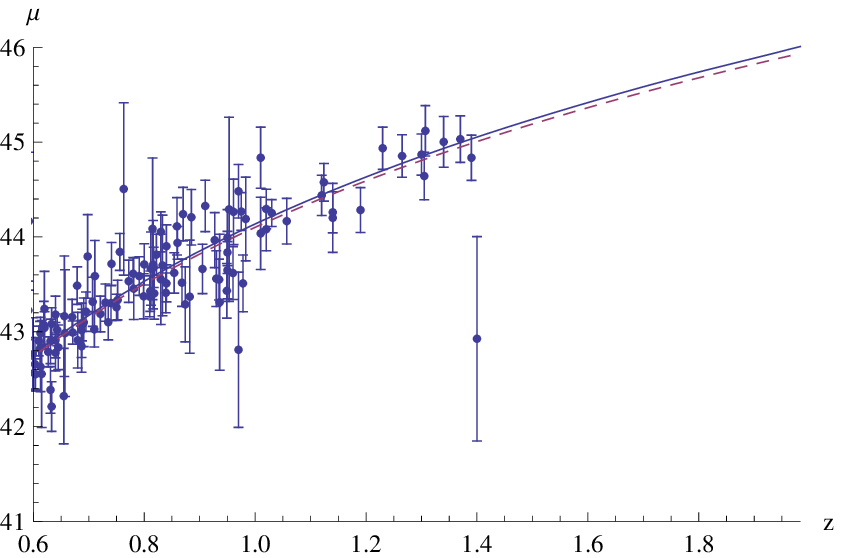}
}
\end{minipage}%\\
\caption{\label{figm3}Comparison with SNe Ia observational data. Solid line corresponds to obtained
numerical solution and dashed line corresponds to $\mathrm{\Lambda CDM}$-model.}
\end{figure}

Calculation $\chi^2$ defined by (\ref{chiall}) in the same grid in parameter region $0.00009\le
\varepsilon\le 0.0002$, $0.12\le\tilde{b}\le 0.2$, $0.66\le \Omega_{\Lambda}\le 0.74$ we find
approximate best fit parameters for this model: $\varepsilon \approx 0.00019$, $\tilde{b}\approx
0.14$, $\Omega_{\Lambda}\approx 0.72$ and $\bar{\mu}_0=18.799$ ($\chi^2 = 542.9$,
$\chi^2/\mathrm{dof}=0.983$). Obtained values of parameters correspond to
$69.9\,\mathrm{km}/(\mathrm{sec}\cdot\mathrm{Mpc})$ for Hubble constant at present epoch,
$\alpha^{-1}=5.11\cdot 10^{-30}\,\mathrm{g}/\mathrm{cm}^3$
 and $\omega_\mathrm{B}=0.0198$ which is in accordance with
data on ${}^3$He-abundance, and lies in $2\sigma$ interval for D-abundance. Solution presented in
Figures~\ref{figm1}--\ref{figm3} corresponds to this set of indefinite parameters. Comparison of
the dependence of the distance modulus $\mu$ as a function of redshift $z$ for obtained numerical
solution with that in $\mathrm{\Lambda CDM}$-model and Union2 observation data is presented in
Figure~\ref{figm3}.

\section{Statefinder diagnostics}

It was shown in a number of papers \cite{statefinder2,statefinder3} that so-called statefinder
diagnostics proposed by Sahni, Saini, Starobinsky and Alam \cite{statefinder1} allows to
effectively discriminate between different models of dark matter and dark energy using the data
from future SNAP-type satellite missions \cite{snap}.

\begin{figure}[t]
\begin{minipage}{0.48\textwidth}
\centering{
 \includegraphics[width=\linewidth]{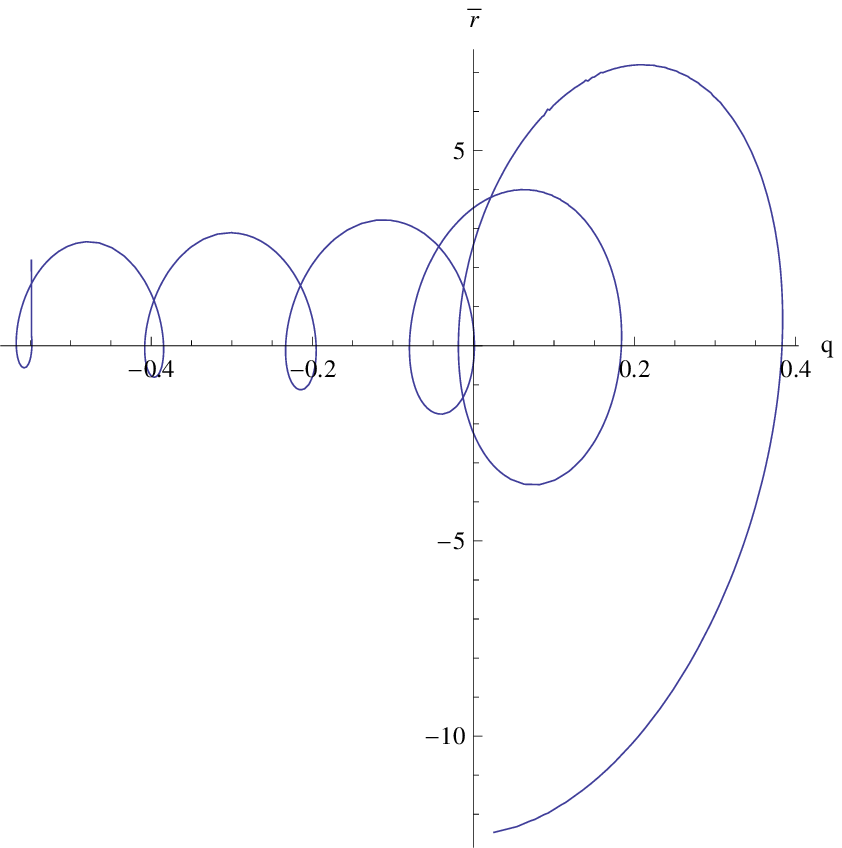}
}
\end{minipage}\, \hfill\,
\begin{minipage}{0.48\textwidth}
\centering{
 \includegraphics[width=\linewidth]{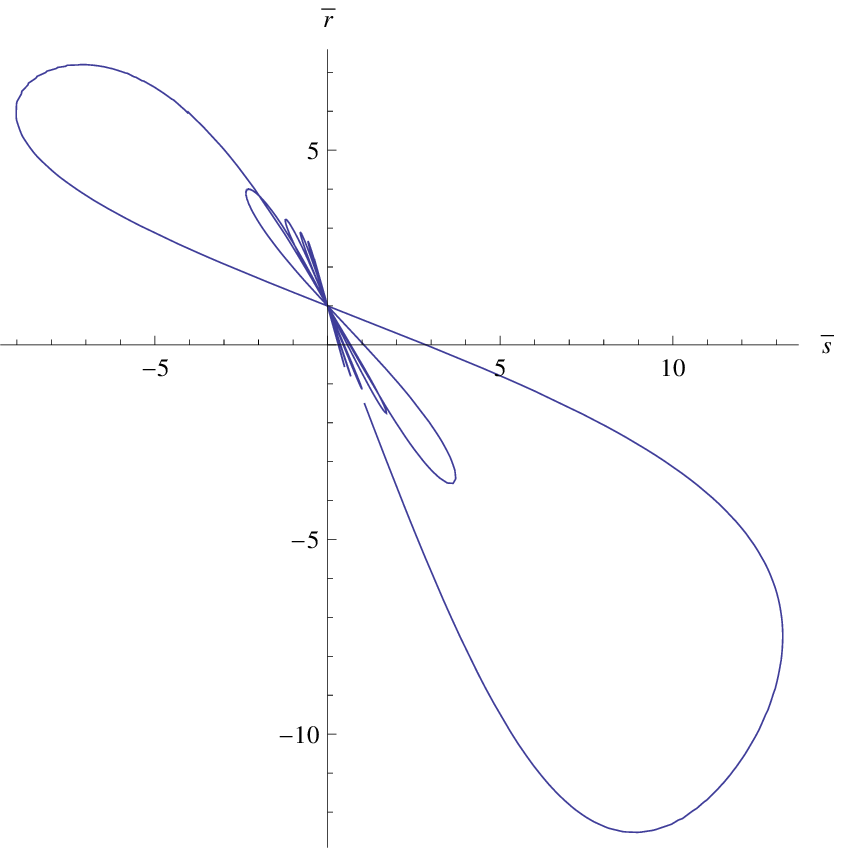}
}
\end{minipage}%\\
\caption{\label{figstatefin1}Evolutionary trajectories of statefinder pair}
\end{figure}

Statefinder diagnostics was applied to cosmology based on dynamic scalar torsion sector of PGTG
\cite{statefindertorsion} and it was found that some characteristics  of the evolution of
statefinder parameters
\begin{equation}\label{statefinderparam}
 \bar{r}=R'''/RH^3 \qquad \text{and} \qquad \bar{s}=(\bar{r}-1)/3(q-1/2)
\end{equation}
can be distinguished from that for other cosmological models.\footnote{The statefinder parameter
$\bar{r}$ is also known as jerk $j$ \cite{visser}.}  The evolutionary trajectories of the
statefinder pair ($\bar{r}$, $\bar{s}$) for numerical solution obtained in Section~\ref{nintsec}
are shown in the Figure~\ref{figstatefin1} in the ($q$, $\bar{r}$) and ($\bar{s}$, $\bar{r}$)
planes. This Figure demonstrates that statefinder diagnostics allows to distinguish cosmology in
considered sector of PGTG from cosmology based on scalar torsion sector of PGTG
\cite{statefindertorsion} and other cosmological models, that allows in principle to differentiate
the considered cosmological models from others using future SNAP-type satellite missions. It is
necessary to note complicated and oscillatory behaviour of statefinder pair which is result of
oscillating behaviour of the deceleration parameter (see Figure~\ref{figm2}). As result the
successful comparison observation data from planned SNAP-missions may require particular procedure
for observation data processing.

\begin{figure}[t]
\begin{minipage}{0.48\textwidth}
\centering{
 \includegraphics[width=\linewidth]{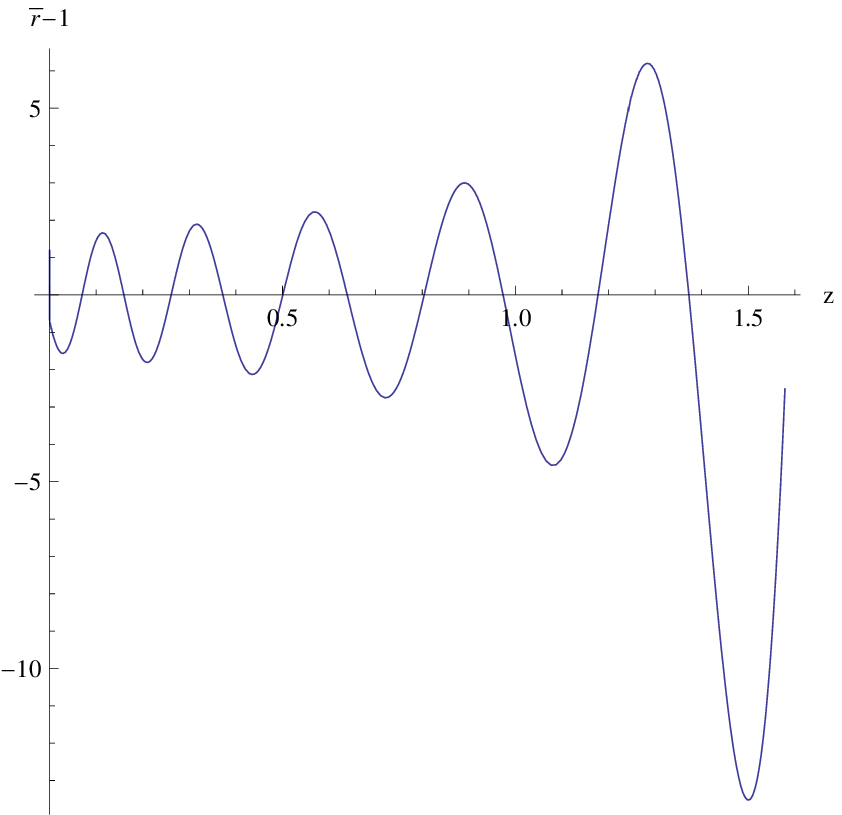}
}
\end{minipage}\, \hfill\,
\begin{minipage}{0.48\textwidth}
\centering{
 \includegraphics[width=\linewidth]{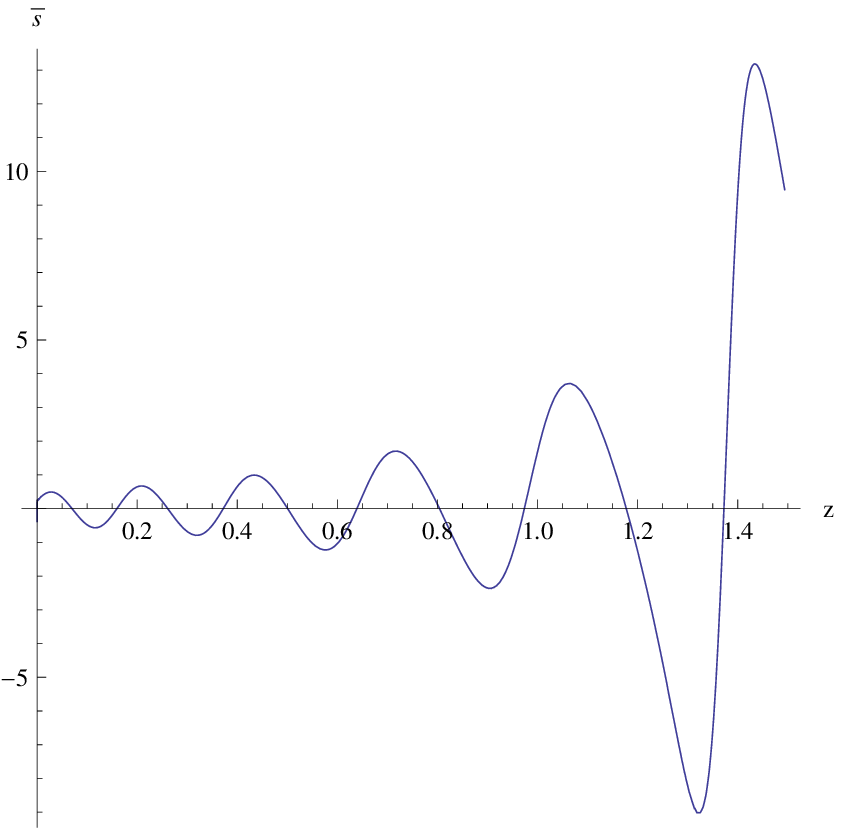}
}
\end{minipage}%\\
\caption{\label{figstatefin2}Statefinder pair as function of redshift $z$.}
\end{figure}

One of the possible procedure may consist in calculation of the statefinder pair ($\bar{r}$,
$\bar{s}$) based on data from specially selected intervals ($z_i$, $z_{i+1}$). For example,
averaging of the statefinder pair for the obtained numerical solution in the range $0.069\le z\le
0.26$ gives $\left<\bar{r}\right>=0.88$ and $\left<\bar{s}\right>=0.057$ which is close to that for
$\mathrm{\Lambda CDM}$-model $(\bar{r},\, \bar{s})_{\mathrm{\Lambda CDM}}=(1,\,0)$, but averaging
over intervals  $0.069\le z\le 0.16$ and $0.16\le z\le 0.26$ gives $(\left<\bar{r}\right>,\,
\left<\bar{s}\right>)$ equal to $(2.05,\, -0.36)$ and $(-0.15,\, 0.43)$ correspondingly. Thus,
averaging over twice smaller interval demonstrates different values and oscillations near the point
$(1,\, 0))$. This feature in the behaviour of the state finder pair can be used as a crucial test
of the considered model in the planned SNAP-type satellite missions, but comprehensive analysis of
possible tests of such type including the procedure of $z_i$ determination is out of this paper.

\section*{Conclusion}

As follows from our analysis, homogeneous isotropic models  built in the
framework of the Poincar\'e gauge theory of gravity and filled by ideal fluid
can have stable solutions with de Sitter asymptotics, if certain restrictions
on indefinite parameters of the gravitational Lagrangian are imposed. Obtained
model demonstrates accelerated expansion at the late time approximation and
does not include dark energy for which physical nature is still unknown.
Contribution of dark matter to the matter energy density is a free parameter of
considered model and it can be either vanishing or non-vanishing. In the case
of vanishing dark matter only baryonic matter with dust equation of state
contributes to the total energy density in considered model.

Correspondence with SNe Ia observation data and BBN predictions are analyzed, where restricted set
of indefinite parameters was considered and special procedure for determination of an initial
conditions was used. Using this procedure best fit values of indefinite parameters and initial
conditions are found. Obtained numerical solution was shown to be in a good correspondence with
$\mathrm{\Lambda CDM}$-model as well as SNe Ia observation data and in an accordance with data on
${}^3$He- and D-abundance.

It was shown, that the trajectories of the statefinder pair for obtained solution demonstrate
behaviour different from that in dynamic scalar torsion sector of PGTG and other cosmological
models, that allows in principle to discriminate the considered cosmological models from others.
Special feature in the behaviour of this trajectories is noticed allowing to test considered model
using data from future SNAP-type satellite missions.

\end{document}